\documentclass[a4paper,12pt]{article}

\makeatletter
 
  \@addtoreset{equation}{section}
 \makeatother
\usepackage{amsfonts}
\usepackage{amssymb}
\usepackage{mathrsfs}
\usepackage{amsmath,amsthm}
\usepackage{amsmath}

\usepackage{graphicx}
\usepackage{color}
\usepackage{indentfirst}

\usepackage{url}

\begin{document}

\newtheorem{theorem}{Theorem}
\newtheorem{lemma}{Lemma}
\newtheorem{prop}{Proposition}
\newtheorem{cor}{Corollary}
\theoremstyle{definition}
\newtheorem{defn}{Definition}
\newtheorem{remark}{Remark}
\newtheorem{step}{Step}

\newcommand{\Cov}{\mathop {\rm Cov}}
\newcommand{\Var}{\mathop {\rm Var}}
\newcommand{\E}{\mathop {\rm E}}
\newcommand{\const }{\mathop {\rm const }}
\everymath {\displaystyle}

\newcommand{\ruby}[2]{
\leavevmode
\setbox0=\hbox{#1}
\setbox1=\hbox{\tiny #2}
\ifdim\wd0>\wd1 \dimen0=\wd0 \else \dimen0=\wd1 \fi
\hbox{
\kanjiskip=0pt plus 2fil
\xkanjiskip=0pt plus 2fil
\vbox{
\hbox to \dimen0{
\small \hfil#2\hfil}
\nointerlineskip
\hbox to \dimen0{\mathstrut\hfil#1\hfil}}}}

\def\qedsymbol{$\blacksquare$}
\renewcommand{\thefootnote }{\fnsymbol{footnote}}

\renewcommand{\refname }{References}

\title{VWAP Execution as an Optimal Strategy
\footnote{This paper is a long version of the paper published in {\it JSIAM Letters}, Vol. 7 (2015), 33--36. DOI: 10.1007/s00780-014-0232-0.}}

\author{Takashi Kato
\footnote{Division of Mathematical Science for Social Systems, 
              Graduate School of Engineering Science, 
              Osaka University, 
              1-3, Machikaneyama-cho, Toyonaka, Osaka 560-8531, Japan, 
E-mail: \texttt{kato@sigmath.es.osaka-u.ac.jp}}
}

\date{First Version: August 26th, 2014\\
This Version: January 31st, 2017}

\maketitle

\begin{abstract}
The volume weighted average price (VWAP) execution strategy is
well known and widely used in practice.
In this study, we explicitly introduce a trading volume process into the Almgren--Chriss model,
which is a standard model for optimal execution. We then
show that the VWAP strategy is the optimal execution strategy for a risk-neutral trader.
Moreover, we examine the case of a risk-averse trader and
derive the first-order asymptotic expansion of the optimal strategy
for a mean-variance optimization problem.
\end{abstract}

\everymath {\displaystyle}

\section{Introduction}
Recently, researchers in the field of mathematical finance have increasingly begun studying the optimal execution problem. Bertsimas and Lo \cite {Bertsimas-Lo} and
Almgren and Chriss \cite {Almgren-Chriss} are
the classic and standard studies in the field, while
Gatheral and Schied \cite {Gatheral-Schied} provide a survey
of dynamic models that address execution problems.

When studying execution problems,
we 
should take care of the market impact (MI),
which is the effect that a trader's investment behavior has on security prices.
As pointed out by \cite {Kato_OU},
the price recovery effect is efficient as another property
that affects the trader's execution schedule.
The price recovery effect is also recognized as the resilience of the MI or transient MI,
and several studies have proposed optimal execution models
with MI functions and resilience functions 
(see \cite {Gatheral-Schied} and the references therein).

Furthermore, trading volume (turnover)
is another important factor in execution problems.
Trading volume is a representative index of financial market activity.
If trading volume is high, the security is highly liquid, and
a trader can easily liquidate shares of the security.
As an execution strategy that exploits trading volume,
the volume weighted average price (VWAP) strategy is
well known and widely used in practice \cite {Madhavan}.
The execution speed of the VWAP strategy is
proportional to the trading volume of the relevant security.
Although the VWAP strategy is a standard execution strategy,
it remains unclear why it is effective in terms of optimal execution theory.

The purpose of this study is to investigate
whether the VWAP strategy is in fact optimal
in an execution problem
equipped with explicitly defined trading volume processes.
Here, we introduce the generalized Almgren--Chriss (AC) model,
in which the temporary MI function is affected by the trading volume.
Then, we show that the VWAP strategy is optimal when the trader is risk neutral.
Moreover, we study the case of a risk-averse trader and deterministic trading volume processes 
and present the second-order linear ordinal differential equation (ODE),
in which the solution is the optimal strategy of the corresponding
mean-variance optimization problem.
We also provide the first-order asymptotic expansion for the optimal strategy.
Finally, we study the mean variance optimization problem when the trading volume process 
follows the geometric Brownian motion.

\section{VWAP and VWAP strategies}\label{sec_VWAP}
In this section, we briefly introduce VWAP and define VWAP execution strategies.
First, $S_t$ denotes a security price at time $t\in [0, T]$, where $T > 0$ is a time horizon.
Mathematically, $(S_t)_t$ is regarded as a continuous-time stochastic process.
The strict definition of $(S_t)_t$ is omitted in this section, but is provided in the next section.

Then, $V_t$ denotes a cumulative trading volume process on the time interval $[0, T]$.
We assume that $V_t$ is continuously differentiable and strictly increasing. That is,
there is a positive continuous process $(v_t)_t$, such that
\begin{eqnarray}\label{def_Vt}
V_t = \int ^t_0v_rdr, \ \ t\in [0, T], \ \ \mbox {a.s.}
\end{eqnarray}
Then, the VWAP of the security at $t = T$ is defined as
\begin{eqnarray}\label{market_VWAP}
S^\mathrm {VWAP}_T = \frac{\int ^T_0S_tv_tdt}{\int ^T_0v_tdt} = \frac{1}{V_T}\int ^T_0S_tv_tdt.
\end{eqnarray}

Next, we define the trader's own VWAP.
We consider a single trader who has a large number of shares, $\Phi > 0$, at initial time $t = 0$,
which he/she tries to liquidate until $t = T$.
A trader's execution strategy is denoted by ${\boldsymbol \zeta } = (\zeta _t)_t$, where
$\zeta _t\geq 0$ is the execution (selling) speed at time $t$.
We assume the following ``sell-off condition'' \cite {Kato_FS}:
\begin{eqnarray}\label{SO}
\int ^T_0\zeta _tdt = \Phi .
\end{eqnarray}
This condition implies that the trader must sell all securities held until the time horizon.
The trader's VWAP is defined by
\begin{eqnarray*}
S^\mathrm {vwap}_T({\boldsymbol \zeta }) =
\frac{\int ^T_0S_t\zeta _tdt}{\int ^T_0\zeta _tdt}.
\end{eqnarray*}

Then, the VWAP execution strategy are defined as follows.

\begin{defn}
${\boldsymbol \zeta } = (\zeta _t)_t$ is called a VWAP strategy
if $\zeta _t = \gamma v_t$, $t\in [0, T]$ a.s., for some $\gamma > 0$.
\end{defn}
Note that $\gamma $ is called an involvement ratio.
It is immediately shown that if ${\boldsymbol \zeta }$ is a VWAP strategy, then
$S^\mathrm {vwap}_T({\boldsymbol \zeta })$ coincides with $S^\mathrm {VWAP}_T$ a.s.
To sell $\Phi $ shares of the security using the VWAP strategy,
$\gamma $ is set so that ${\boldsymbol \zeta }$ satisfies (\ref {SO}).
That is,
\begin{eqnarray*}
\Phi = \int ^T_0\gamma v_tdt = \gamma V_T,
\end{eqnarray*}
which implies that $\gamma = \Phi / V_T$.
However, such a strategy cannot be executed in practice,
because the value of $V_T$ is unobservable before the time horizon, $T$.

The difference between two VWAPs, $S^\mathrm {vwap}_T({\boldsymbol \zeta })$ and $S^\mathrm {VWAP}_T$,
is called a VWAP slippage. Minimizing VWAP slippage problems is studied by \cite {Frei-Westray, Konishi}.
See also \cite {Gueant-Royer}. 

\section{The AC model with trading volume}
In this section, we introduce our model of an optimal execution problem.
Our model is based on the AC model, as proposed by \cite {Almgren-Chriss} and
generalized by the authors of \cite {Gatheral-Schied}.

Let $(\Omega , \mathcal {F}, (\mathcal {F}_t)_t, P)$ be a stochastic basis and
let $(S^0_t)_t$ be an $(\mathcal {F}_t)_t$-martingale satisfying
\begin{eqnarray*}
\E [(S^0_T)^2] < \infty .
\end{eqnarray*}
Here, $S^0_t$ is regarded as an unaffected price of the security at time $t$. In other words,
it is the security price when there is no MI.
For simplicity, we assume that $(S^0_t)_t$ is independent of $(v_t)_t$. 
Note that this assumption is not required when we consider static or adaptive optimization. 

The execution strategy ${\boldsymbol \zeta } = (\zeta _t)_t$ is referred to as admissible 
if ${\boldsymbol \zeta } = (\zeta _t)_t$ is 
$(\mathcal {F}_t)_t$-adapted and
satisfies the sell-off condition (\ref {SO}).
The set of admissible strategies is denoted by $\mathcal {A}(\Phi )$.
Under the given admissible strategy, ${\boldsymbol \zeta }\in \mathcal {A}(\Phi )$,
the security price, $S_t$, is defined as follows:
\begin{eqnarray}\label{def_S}
S_t = S^0_t - \int ^t_0g(\zeta _r)dr - \tilde{g}(v_t, \zeta _t),
\end{eqnarray}
where $g$ (resp., $\tilde{g}$) is a permanent (resp., temporary) MI function and
$(v_t)_t$ is a positive $(\mathcal {F}_t)_t$-adapted process describing the instantaneous trading volume.

In this study, $g$ is always assumed to be a linear function:
\begin{eqnarray*}
g(\zeta ) = \kappa \zeta,
\end{eqnarray*}
for some $\kappa > 0$.
This assumption is necessary given the view of an absence of price manipulation. See \cite {Gatheral-Schied} for more detail.

The temporary MI function, $\tilde{g}$,
depends on the execution strategy, $\zeta _t$, and the trading volume, $v_t$.
It is natural that the temporary MI is decreasing as the trading volume increases,
because a large trading volume implies high market liquidity.
Therefore, we adopt the following form as the function $\tilde{g}$:
\begin{eqnarray}\label{tilde_g}
\tilde{g}(v, \zeta ) = \frac{\tilde{\kappa }\zeta }{v},
\end{eqnarray}
where $\tilde{\kappa } > 0$.

\begin{remark}We can generalize the form of $\tilde{g}$ as
\begin{eqnarray}\label{tilde_g_general}
\tilde{g}(v, \zeta ) = k(v)\zeta ^\alpha,
\end{eqnarray}
where
$\alpha > 0$ and
$k$ is a positive continuous function.
See also Remark \ref {rem_general} below.
\end{remark}

Next, we define our objective function.
For a given ${\boldsymbol \zeta }$,
an implementation shortfall (IS) cost is defined as
\begin{eqnarray}\label{def_cost}
\mathcal {C}({\boldsymbol \zeta }) = S_0\Phi - \int ^T_0S_t\zeta _tdt.
\end{eqnarray}
Substituting (\ref {def_S}) and applying the
integration by parts formula, we can rewrite (\ref {def_cost}) as
\begin{eqnarray*}
\mathcal {C}({\boldsymbol \zeta }) =
\frac{\kappa \Phi ^2}{2} -
\int ^T_0\varphi _tdS^0_t + \tilde{\kappa }\int ^T_0\frac{\zeta ^2_t}{v_t}dt,
\end{eqnarray*}
where
\begin{eqnarray}\label{def_varphi}
\varphi _t = \Phi - \int ^t_0\zeta _rdr
\end{eqnarray}
denotes
the remaining shares of the security held at time $t$.

We define the following three classes of
admissible strategies:
\begin{eqnarray*}
\mathcal {A}^\mathrm {ant}(\Phi ) &=&
\{ {\boldsymbol \zeta }\ ; \ (\hat{\mathcal {G}}_t)_t\mbox{-adapted, satisfying (\ref {SO})} \} , \\
\mathcal {A}^\mathrm {adap}(\Phi ) &=&
\{ {\boldsymbol \zeta }\in \mathcal {A}(\Phi )\ ; \ (\mathcal {G}_t)_t\mbox{-adapted} \} , \\
\mathcal {A}^\mathrm {stat}(\Phi ) &=&
\{ {\boldsymbol \zeta }\in \mathcal {A}(\Phi )\ ; \ 
\mbox {deterministic} \} , 
\end{eqnarray*}
where $\hat{\mathcal {G}}_t = \mathcal {H}_t\vee \mathcal {G}_T$ and 
$(\mathcal {G}_t)_t$ (resp., $(\mathcal {H}_t)_t$) is a filtration generated by $(v_t)_t$ (resp., $(S^0_t)_t$). 
(Note that ${\boldsymbol \zeta }\in \mathcal {A}^\mathrm {ant}(\Phi )$ is not assumed to be $(\mathcal {F}_t)_t$-adapted.) 

Strategy ${\boldsymbol \zeta }\in \mathcal {A}^\mathrm {ant}(\Phi )$
is called an anticipating strategy.
In this case, the trader knows the value of $V_T$ (the cumulative trading volume defined by (\ref {def_Vt}))
at the initial time. Therefore, he/she can execute the VWAP strategy with the involvement ratio $\Phi / V_T$. 
This case is unrealistic, 
but the optimal anticipating strategy gives us
a benchmark for execution strategies.

Strategy ${\boldsymbol \zeta }\in \mathcal {A}^\mathrm {adap}(\Phi )$ is called an adaptive strategy.
It is natural to search for an optimal strategy from within the adaptive strategies,
but the problem may become difficult.

Strategy ${\boldsymbol \zeta }\in \mathcal {A}^\mathrm {stat}(\Phi )$ is called a static strategy.
In trading practice, a VWAP strategy is often classified as a static (deterministic) strategy.
In this case, an estimated cumulative trading volume 
is used as a substitute for $V_T$.

\subsection{Risk-neutral case}
In this subsection, we examine the case in which the trader is risk neutral.
Our purpose is to minimize the expected IS cost,
\begin{eqnarray}\label{exp_cost}
\E [\mathcal {C}({\boldsymbol \zeta })] = \frac{\kappa \Phi ^2}{2} +
\tilde{\kappa }\E \left[\int ^T_0\frac{\zeta ^2_t}{v_t}dt \right] .
\end{eqnarray}
Here, equality (\ref {exp_cost}) follows from the martingale property of $(S^0_t)_t$.
We can then show the following two theorems by applying the Jensen inequality
and appropriate changes of variables (see Section \ref {sec_proofs} for the proofs).

\begin{theorem}\label{th_ant}
$\hat{\zeta }_t = v_t\Phi /V_T$
is the optimal anticipating strategy for the problem
$\inf _{{\boldsymbol \zeta }\in \mathcal {A}^\mathrm {ant}(\Phi )}\E [\mathcal {C}({\boldsymbol \zeta })]$.
\end{theorem}

\begin{theorem}\label{th_stat}
Set $u_t = 1/\E [1/v_t]$ and assume that $u_t$ is finite and continuous in $t$.
Then, $\tilde{\zeta }_t = u_t\Phi /U_T$
is the optimal static strategy for the problem
$\inf _{{\boldsymbol \zeta }\in \mathcal {A}^\mathrm {stat}(\Phi )}\E [\mathcal {C}({\boldsymbol \zeta })]$,
where 
$U_T = \int ^T_0u_tdt$.
\end{theorem}

The strategy $\hat{{\boldsymbol \zeta }} = (\hat{\zeta }_t)_t$ is the VWAP strategy.
Therefore, Theorem \ref {th_ant} gives the typical case
in which the VWAP strategy is optimal for the execution problem.
Note again that the ``exact'' VWAP strategy, $\hat{{\boldsymbol \zeta }}$, is
not executable without observing the future value of the cumulative trading volume, $V_T$.
However, it provides a benchmark strategy for the optimal execution problem.

Theorem \ref {th_stat} tells us how to mimic the VWAP strategy, $\hat{{\boldsymbol \zeta }}$.
The expected VWAP strategy $\tilde {{\boldsymbol \zeta }} = (\tilde{\zeta }_t)_t$ is optimal
in the class of static strategies, and the expectation $u_t$ of $v_t$ is
calculated in the sense of a harmonic mean.
In practice, the estimated value of the trading volume based on historical data
is often applied to replicate the VWAP strategy.
The assertion of Theorem \ref {th_stat} is consistent with such a situation.

\begin{remark}\label{rem_general}
We can also show the above two theorems when
$\tilde{g}$ has a more general form, as in $(\ref {tilde_g_general})$.
In this case, $\hat{\zeta }_t$ is given as $\hat{\zeta }_t = \tilde{v}_t\Phi / \tilde{V}_T$, where
\begin{eqnarray*}
\tilde{v}_t = \frac{1}{k(v_t)^{1/\alpha }}, \ \ \tilde{V}_T = \int ^T_0\tilde{v}_tdt.
\end{eqnarray*}
We refer to $(\hat{\zeta }_t)_t$ as a ``twisted'' VWAP strategy.
Similarly, $\tilde{\zeta }_t$ is changed to $\tilde{\zeta }_t = \tilde{u}_t\Phi / \tilde{U}_T$, where
\begin{eqnarray*}
\tilde{u}_t = \frac{1}{\E [k(v_t)]^{1/\alpha }}, \ \ \tilde{U}_T = \int ^T_0\tilde{u}_tdt.
\end{eqnarray*}
\end{remark}

Finally, we consider the adaptive case.
Here, we only consider the special case in which $(v_t)_t$ follows a geometric Brownian motion:
\begin{eqnarray}\label{SDE_v}
dv_t = v_t(\mu dt + \sigma dB_t), \ \ t\geq 0, \ \ v_0 > 0, 
\end{eqnarray}
where $\mu \in \Bbb {R}$, $\sigma > 0$, and $(B_t)_t$ is a one-dimensional
$(\mathcal {F}_t)_t$-Brownian motion.
Then, we obtain the following theorem using an adequate verification argument of dynamic control theory 
(see Theorems 1--2 in \cite {Kato_AC} for the details). 

\begin{theorem}\label{th_adap}
Assume $\E [U_T] < \infty $.
Then $\tilde{\zeta }_t$, defined in Theorem \ref {th_stat},
is the optimal adaptive strategy for the problem
$\inf _{{\boldsymbol \zeta }\in \mathcal {A}^\mathrm {adap}(\Phi )}\E [\mathcal {C}({\boldsymbol \zeta })]$.
\end{theorem}

This theorem implies that the optimal adaptive strategy coincides with
the expected VWAP strategy, which is also optimal among the static strategies.

\begin{remark}\label{rem_TWAP}
When $v_t$ is a constant,
the optimal strategy is selling with constant speed (i.e., $\hat{\zeta }_t = \tilde{\zeta }_t = \Phi / T$).
This strategy is called a time weighted average price (TWAP) strategy.
In \cite {Kato_FS, Kato_JSIAM},
we find similar examples such that the TWAP strategy is
the optimal strategy for a risk-neutral trader
when the permanent MI function is non-linear.
In particular, Theorem 7.4 in \cite {Kato_JSIAM} gives the example of an
``S-shaped'' MI function, which is often observed in practice.
\end{remark}

\subsection{Mean-variance optimization}
Next, we study the following static mean-variance optimization problem:
\begin{eqnarray}\label{problem_MV}
\inf _{{\boldsymbol \zeta }\in \mathcal {A}^\mathrm {stat}(\Phi )}
\mathrm {MV}^\lambda (\mathcal {C}({\boldsymbol \zeta })),
\end{eqnarray}
where $\lambda \geq 0$ 
and
\begin{eqnarray*}
\mathrm {MV}^\lambda (\mathcal {C}({\boldsymbol \zeta })) =
\mathrm {E}[\mathcal {C}({\boldsymbol \zeta })] + \lambda
\mathrm {Var}(\mathcal {C}({\boldsymbol \zeta })).
\end{eqnarray*}
The parameter $\lambda $ denotes the measure of the trader's risk aversion.
When $\lambda = 0$, the above problem is equivalent to that of the previous subsection.

In this subsection, we assume the following:
\begin{eqnarray*}
S^0_t = S^0_0 + \tilde{\sigma }\tilde{B}_t, 
\end{eqnarray*}
where $\tilde{\sigma } > 0$ and $(\tilde{B}_t)_t$ is 
a one dimensional $(\mathcal {F}_t)_t$-Brownian motion.

\subsubsection{Deterministic turnover}\label{subsubsec_deterministic}

First we consider the case where $(v_t)_t$ is deterministic. 
We assume that 
$v_t \in C^1((0, T))$ and $v_t \geq \delta $, $t\in [0, T]$ for some $\delta > 0$ 
($v_0$ and $v_T$ may diverge).

By a straightforward calculation, we see that
\begin{eqnarray*}
\mathrm {MV}^\lambda (\mathcal {C}({\boldsymbol \zeta })) =
\frac{\kappa \Phi ^2}{2} + \tilde{\kappa }f^\lambda (\Phi ; {\boldsymbol \zeta }),
\end{eqnarray*}
for ${\boldsymbol \zeta }\in \mathcal {A}^\mathrm {stat}(\Phi )$, where
\begin{eqnarray*}
f^\lambda (\Phi  ; {\boldsymbol \zeta }) = 
\int ^T_0\left\{ \frac{\tilde{\sigma }^2\lambda }{\tilde{\kappa }}\varphi _t^2 + \frac{\zeta ^2_t}{v_t} \right\} dt
\end{eqnarray*}
and $(\varphi _t)_t$ is defined as in (\ref {def_varphi}).
Therefore, our optimization problem is equivalent to the following variational problem:
\begin{eqnarray}\label{variational_problem}
\inf _{{\boldsymbol \varphi }}\int ^T_0F(t, \varphi _t, \dot{\varphi }_t)dt,
\end{eqnarray}
subject to $\varphi _0 = \Phi $ and $\varphi _T = 0$,
where ${\boldsymbol \varphi } = (\varphi _t)_{t\in [0, T]}$ is absolutely continuous and
\begin{eqnarray*}
F(t, \varphi , \zeta ) =
\frac{\tilde{\sigma }^2\lambda }{\tilde{\kappa }}\varphi ^2 + \frac{\zeta ^2}{v_t}.
\end{eqnarray*}

From the standard theory of variational analysis (see Section 7.5 of \cite {Luenberger} for instance),
we can show the following theorem.

\begin{theorem}
There is a unique optimizer 
to $(\ref {variational_problem})$
satisfying the following second-order linear ODE with variable coefficients:
\begin{eqnarray}\label{second_ODE}
\ddot{\varphi }_t - a_t\dot{\varphi }_t - \frac{\tilde{\sigma }^2\lambda }{\tilde{\kappa }}\varphi _t = 0, 
\end{eqnarray}
where 
\begin{eqnarray*}
a_t = \frac{\dot {v}_t}{v_t} = \frac{d}{dt}\log v_t. 
\end{eqnarray*}
\end{theorem}

The above theorem implies that we can find the optimizer by
solving the ODE (\ref {second_ODE}) with boundary conditions $\varphi _0 = \Phi $ and $\varphi _T = 0$,
and the optimal strategy of the problem (\ref {problem_MV}) is obtained as $\zeta _t = -\dot{\varphi }_t$.

\begin{remark}
If $v_t$ is a constant,
the optimal strategy is explicitly obtained in \cite {Almgren-Chriss} as
\begin{eqnarray*}
\zeta ^\lambda _t =
\frac{\cosh (\gamma _\lambda (T - t))}{\sinh (\gamma _\lambda T)}\gamma _\lambda \Phi ,
\end{eqnarray*}
where $\gamma _\lambda = \tilde{\sigma }^2\lambda /\tilde{\kappa }$.
The corresponding process, $(\varphi _t)_t$, is given by
\begin{eqnarray}\label{sol_AC}
\frac{\sinh (\gamma _\lambda (T - t))}{\sinh (\gamma _\lambda T)}\Phi .
\end{eqnarray}
In this case,
$(\ref {second_ODE})$ becomes the ODE with constant coefficients.
Then the solution can be explicitly solved and coincides with $(\ref {sol_AC})$.
\end{remark}

Here, we formally derive the first-order asymptotic expansion
of the optimal strategy of (\ref {problem_MV}) with small $\lambda $.
Let ${\boldsymbol \zeta }^\lambda = (\zeta ^\lambda _t)_t$ be the optimal strategy of (\ref {problem_MV}) and
put $\varphi ^\lambda _t = \Phi - \int ^t_0\zeta ^\lambda _rdr$.
We already know that when $\lambda = 0$, it holds that
\begin{eqnarray*}
\zeta ^0_t = \frac{v_t\Phi }{V_T}, \ \ \varphi ^0_t = \left( 1 - \frac{V_t}{V_T}\right) \Phi .
\end{eqnarray*}
Define $\tilde{\varphi }^\lambda _t = (\varphi ^\lambda _t - \varphi ^0_t)/\lambda $.
Then we can easily check that $(\tilde{\varphi }^\lambda _t)_t$ is a solution to
\begin{eqnarray*}
\ddot{\tilde{\varphi }}^\lambda _t - a_t\dot{\tilde{\varphi }}^\lambda _t -
v_t\left (\varphi ^0_t + \frac{\tilde{\sigma }^2\lambda }{\tilde{\kappa }}\tilde{\varphi }^\lambda _t\right ) = 0,
\end{eqnarray*}
with boundary conditions $\tilde{\varphi }^\lambda _0 = \tilde{\varphi }^\lambda _T = 0$.
Letting $\lambda \rightarrow 0$, we get
\begin{eqnarray*}
\ddot{\tilde{\varphi }}^0_t - a_t\dot{\tilde{\varphi }}^0_t =
v_t\left( 1 - \frac{V_t}{V_T}\right) \Phi  .
\end{eqnarray*}
Hence,
\begin{eqnarray*}
\tilde{\varphi }^0_t =
\Phi \left\{ tV_t - \left( 1 + \frac{V_t}{V_T}\right) \mathcal {V}_t - \frac{1}{V_T}\int ^t_0V^2_rdr + CV_t \right\} ,
\end{eqnarray*}
where
\begin{eqnarray*}
C = \frac{2\mathcal {V}_T}{V_T} - T - \frac{1}{V^2_T}\int ^T_0V^2_rdr, \ \
\mathcal {V}_t = \int ^t_0V_rdr.
\end{eqnarray*}
Now we get the following asymptotic expansion formula
around the 
VWAP strategy:
\begin{eqnarray*}
\zeta ^\lambda _t = \frac{v_t\Phi }{V_T} + \lambda \tilde{\zeta }^0_t + o(\lambda ), \ \ \lambda \rightarrow 0,
\end{eqnarray*}
where $\tilde{\zeta }^0_t = -\dot{\tilde{\varphi }}_t$ is given by
\begin{eqnarray*}
\tilde{\zeta }^0_t =
\Phi v_t\left\{ T - t - \frac{\mathcal {V}_T - \mathcal {V}_t}{V_T} -
\frac{1}{V_T}(\mathcal {V}_T - \int ^T_0V^2_rdr)\right\} .
\end{eqnarray*}

\begin{remark}
In fact, we can also solve $(\ref {problem_MV})$ numerically 
(even when $(v_t)_t$ is not deterministic) by, for instance,
using the sequential quadratic programming (SQP) algorithm. 
\end{remark}

\begin{figure}[htbp]
\centering
\includegraphics[scale=0.9]{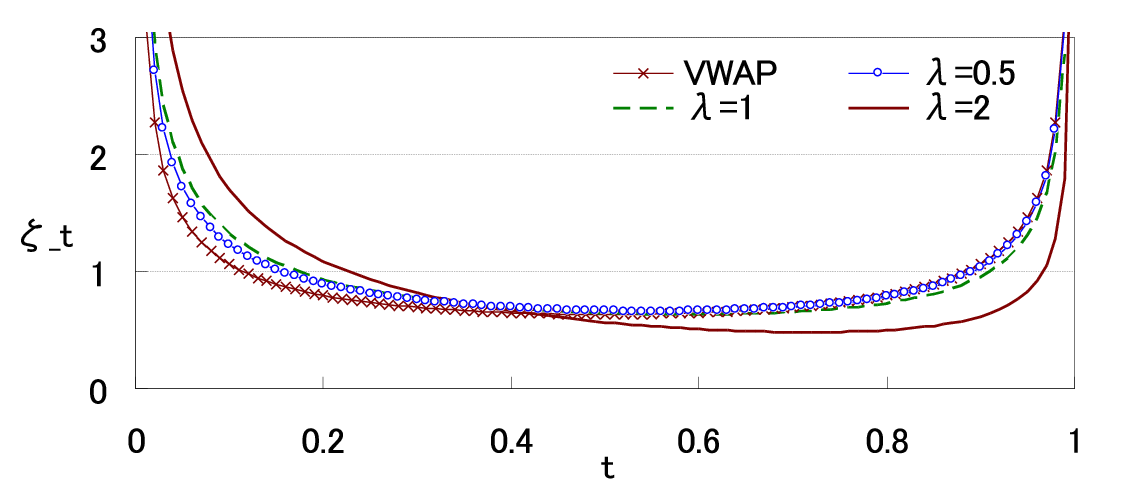}
\caption{The forms of optimal strategies $(\zeta _t)_t$ of (\ref {problem_MV}) with the arcsine cumulative trading volume process; the horizontal axis shows time, $t$; the vertical axis is $\zeta _t$. }
\label{fig_eg}
\end{figure}

Finally, we give a numerical example of an ``arcsine cumulative trading volume process.''
For brevity, set $T = 1$.
Assume that $v_t = 1/(\pi \sqrt{t(1-t)})$.
Then $V_t$ is given by the arcsine function:
$V_t = 2\pi ^{-1}\arcsin \sqrt{t}$.
Note that $v_t \longrightarrow \infty $ as $t\rightarrow 0$ and $t\rightarrow 1$:
this reflects the realistic situation in which the market is active near the opening and closing, but
becomes less active in the continuous session.
Then, we investigate the optimal strategies when setting $\lambda = 0$ (VWAP), $0.5$, $1$, and $2$.
The other parameters are set as $\tilde{\sigma }= 0.1$, $\tilde{\kappa } = 0.02$, and $\Phi = 1$.
The results are shown in Fig. \ref {fig_eg}.
Here, we find that the execution speed with small $t$ becomes larger when $\lambda $ is large.
This is because a large $\lambda $ implies that the trader is risk averse and
wants to quickly liquidate the security 
to avoid volatility risk.

\subsubsection{The Black--Scholes type turnover}

Next we consider the case of stochastic turnover processes. 
We assume that $(v_t)_t$ is given as (\ref {SDE_v}) and that 
\begin{eqnarray*}
d\langle B, \tilde{B} \rangle _t = \rho dt
\end{eqnarray*}
for some $\rho \in [-1, 1]$. 

For each ${\boldsymbol \zeta }\in \mathcal {A}^{\rm stat}(\Phi )$, 
a straightforward calculation gives 
\begin{eqnarray*}
\mathrm {Var}(\mathcal {C}({\boldsymbol \zeta })) &=& 
\E \left[ \left( -\tilde{\sigma }\int ^T_0\varphi _td\tilde{B}_t + 
\tilde {\kappa }\int ^T_0\zeta ^2_tp_tdt \right) ^2\right]\\
&=& \tilde{\sigma }^2\int ^T_0\varphi ^2_tdt - 
2\tilde{\sigma }\tilde{\kappa }\E [M_TA_T] + 
\tilde{\kappa }^2\int ^T_0\int ^T_0\zeta ^2_s\zeta ^2_tC_{s, t}dsdt, 
\end{eqnarray*}
where 
\begin{eqnarray*}
p_t &=& \frac{1}{v_t} - \frac{1}{u_t} = \frac{\exp (-(\mu - \sigma ^2/2)t)}{v_0}
\left\{ e^{-\sigma B_t} - e^{\sigma ^2t/2}\right\} , \\
dM_t &=& \varphi _td\tilde{B}_t, \ \ M_0 = 0, \\
dA_t &=& \zeta ^2_tp_tdt, \ \ A_0 = 0, \\
C_{s, t} &=& \E [p_sp_t] = 
\frac{1}{v_0^2}\exp (-(\mu - \sigma ^2)(t+s))(e^{\sigma ^2(s\wedge t)} - 1). 
\end{eqnarray*}
Since $(M_t)_t$ is a martingale, 
we observe 
\begin{eqnarray*}
\E [M_TA_T] = \E \left [ \int ^T_0M_tdA_t + \int ^T_0A_tdM_t \right ] = 
\frac{1}{v_0}\E \left [ \int ^T_0M_t\zeta ^2_te^{-(\mu - \sigma ^2/2)t - \sigma B_t}dt \right ] 
\end{eqnarray*}
by using Ito's formula. 
Since $(M_t, B_t)$ is the two-dimensional Gaussian random vector with mean zero and 
covariance matrix 
\begin{eqnarray*}
\left(
\begin{array}{cc}
 a_t	& b_t	\\
 b_t	& t
\end{array}
\right) := 
\left(
\begin{array}{cc}
 \int ^t_0\varphi ^2_sds	& \rho \int ^t_0\varphi _sds	\\
 \rho \int ^t_0\varphi _sds	& t
\end{array}
\right) , 
\end{eqnarray*}
we see that 
\begin{eqnarray*}
\E [M_TA_T] = 
\frac{1}{2\pi v_0}\int ^T_0\zeta ^2_te^{-(\mu - \sigma ^2/2)t}
\sqrt{a_t}\int ^\infty _{-\infty }\int ^\infty _{-\infty }
ze^{-\sigma \sqrt{t}(\rho _tz + \sqrt{1 - \rho ^2_t}w)}
e^{-(z^2 + w^2)/2}dzdwdt. 
\end{eqnarray*}
Therefore, we obtain 
\begin{eqnarray*}
\mathrm {MV}^\lambda (\mathcal {C}({\boldsymbol \zeta })) &=& 
\frac{\kappa \Phi ^2}{2} + \tilde{\kappa }\int ^T_0\frac{\zeta ^2_t}{u_t}dt\\
&& + \lambda \Big \{ 
\tilde{\sigma }^2a_T - 
\frac{\tilde{\sigma }\tilde{\kappa }}{\pi v_0}
\int ^T_0\zeta ^2_te^{-(\mu - \sigma ^2/2)t}
\sqrt{a_t}\int _{\Bbb {R}^2}ze^{-\sigma \sqrt{t}(\rho _tz + \sqrt{1 - \rho ^2_t}w)}e^{-(z^2 + w^2)/2}dzdwdt\\
&&\hspace{8mm} + 
\frac{\tilde{\kappa }^2}{v_0^2}\int ^T_0\int ^T_0\zeta ^2_s\zeta ^2_t
\exp (-(\mu - \sigma ^2)(t+s))(e^{\sigma ^2(s\wedge t)} - 1)dsdt \Big \} . 
\end{eqnarray*}
We can numerically solve the optimization problem (\ref {problem_MV}) by using the SQP algorithm and 
the Gaussian quadrature method. 

Now we give some results of the numerical experiment. 
We always set $\sigma = 0.2$, $\tilde{\sigma } = 0.2$, $\tilde{\kappa } = 0.02$, $v_0 = 1$, and $\Phi = 1$. 
$\mu $ is set as $\mu = -\sigma ^2/2$ so that $(v_t)_t$ becomes a martingale. 

\begin{figure}[htbp]
\centering
\includegraphics[scale=0.9]{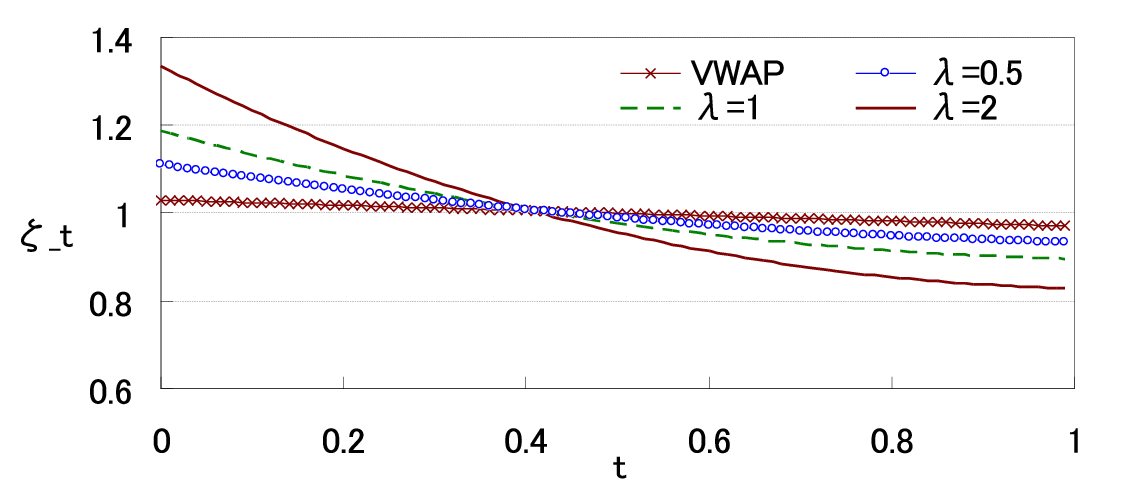}
\caption{The forms of optimal strategies $(\zeta _t)_t$ of (\ref {problem_MV}) with the Black--Scholes type trading volume process with $\rho = 0$; the horizontal axis shows time, $t$; the vertical axis is $\zeta _t$. }
\label{fig_2}
\end{figure}

First we investigate the effect of $\lambda $. 
Figure \ref {fig_2} shows the forms of optimal strategies 
with $\rho = 0$ and $\lambda = 0$ (VWAP), $0.5, 1$, and $2$. 
Similarly to the case of Section \ref {subsubsec_deterministic}, 
we find that the execution speed with small $t$ becomes larger 
together with $\lambda $. 

\begin{figure}[htbp]
\centering
\includegraphics[scale=0.9]{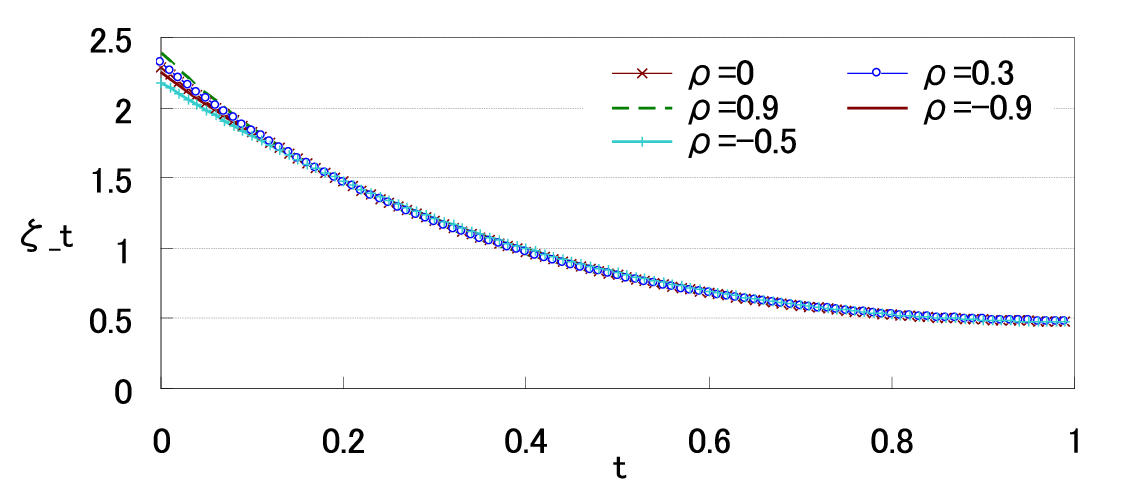}
\caption{The forms of optimal strategies $(\zeta _t)_t$ of (\ref {problem_MV}) with the Black--Scholes type trading volume process with $\lambda = 10$; the horizontal axis shows time, $t$; the vertical axis is $\zeta _t$. }
\label{fig_3}
\end{figure}

Next we study the effect of the correlation parameter $\rho $. 
We fix $\lambda = 10$ and we numerically calculate the optimal strategies with 
$\rho = 0, \pm 0.3$, and $\pm 0.9$. 
The results are in Figure \ref {fig_3}. 
We see that the execution speed with small $t$ also moves the same direction as $\rho $, 
i.e., $\zeta _t$ with small $t$ becomes large when $\rho $ is large. 
However, the effect of $\rho $ is not so large (in this parameter settings.)

\section{Concluding remarks}
In this study, we studied the generalized AC model while considering
trading volume. 
Here, we showed that, with some standard settings, 
the VWAP execution strategy is the optimal strategy for a risk-neutral trader.
When considering adaptive strategies,
we obtained our result only when the trading volume process follows
a geometric Brownian motion.
Studying more general cases of adaptive optimization problems is an area of future research.

In this study, we only considered volume-dependent temporary MI functions.
Therefore, it remains to study the case in which
the permanent MI function depends on the trading volume.

As mentioned in Remark \ref {rem_TWAP},
when the trader is risk neutral,
there are examples other than the AC-based model
in which the TWAP strategy is optimal.
Therefore, it would also be interesting to investigate whether
the optimal strategy is a VWAP strategy
in the model of \cite {Kato_FS, Kato_JSIAM} by explicitly introducing trading volume processes so that
the time parameter is replaced by ``volume time (stochastic clock).''

\appendix 

\section{Proof of Theorems \ref {th_ant}--\ref {th_stat}}\label{sec_proofs}

\noindent 
{\it Proof of Theorem \ref {th_ant}.}

First note that 
\begin{eqnarray*}
\E [\mathcal {C}(\hat {\boldsymbol \zeta })] = 
\left( \frac{\kappa }{2} + \tilde{\kappa }\E \left [\frac{1}{V_T}\right ]\right) \Phi ^2. 
\end{eqnarray*}
On the other hand, for any ${\boldsymbol \zeta }\in \mathcal {A}^\mathrm {ant}(\Phi )$, we see that 
\begin{eqnarray*}
\E [\mathcal {C}({\boldsymbol \zeta })] - \frac{\kappa \Phi ^2}{2} &=& 
\tilde{\kappa }\E \left[ \int ^{V_T}_0\left( \frac{\zeta _{V^{-1}_{\tilde {t}}}}{v_{V^{-1}_{\tilde {t}}}}\right) ^2 d\tilde{t} \right]  \\
&\geq & 
\tilde{\kappa }\E \left[ \frac{1}{V_T}\left( \int ^{V_T}_0\frac{\zeta _{V^{-1}_{\tilde {t}}}}{v_{V^{-1}_{\tilde {t}}}}d\tilde{t}\right) ^2 \right]  = 
\tilde{\kappa }\E \left [\frac{1}{V_T}\right ]\Phi ^2. 
\end{eqnarray*}
due to the Jensen inequality. 
These imply that $\E [\mathcal {C}(\hat {\boldsymbol \zeta })] = \inf _{{\boldsymbol \zeta }\in \mathcal {A}^\mathrm {ant}(\Phi )}\E [\mathcal {C}({\boldsymbol \zeta })]$. \qed 
\\\\

\noindent 
{\it Proof of Theorem \ref {th_stat}.}

The Jensen inequality implies that it holds for any ${\boldsymbol \zeta }\in \mathcal {A}^\mathrm {stat}(\Phi )$ that 
\begin{eqnarray*}
\E [\mathcal {C}({\boldsymbol \zeta })] - \frac{\kappa \Phi ^2}{2} &=& 
\tilde {\kappa }\int ^{U_T}_0\left( \frac{\zeta _{U^{-1}_{\tilde{t}}}}{u_{U^{-1}_{\tilde{t}}}}\right) ^2d\tilde{t}\\
&\geq & 
\frac{\tilde {\kappa }}{U_T}\left( \int ^{U_T}_0\frac{\zeta _{U^{-1}_{\tilde{t}}}}{u_{U^{-1}_{\tilde{t}}}}d\tilde{t}\right) ^2 = \frac{\tilde{\kappa }\Phi ^2}{U_T}. 
\end{eqnarray*}
Obviously it holds that 
$\E [\mathcal {C}(\tilde {\boldsymbol \zeta })] = 
\left( \frac{\kappa }{2} + \frac{\tilde{\kappa }}{U_T}\right) \Phi ^2$, thus we obtain the assertion. \qed 

\section{An Application to an Inequality on Harmonic Mean}

As a consequence of Theorems \ref {th_ant}--\ref {th_stat}, 
we can immediately show a generalized version of Proposition 14.6.3 in \cite {Rao}, 
which implies a relationship between ``the harmonic mean of the expected value'' and 
``the expected value of the harmonic mean.'' 

\begin{theorem}\label{th_harmonic}Let $(X, \mathscr {X}, \mu )$ and $(Y, \mathscr {Y}, \nu )$ be measure spaces and 
let $f : X\times Y\longrightarrow (0, \infty )$ be an $\mathscr {X}\otimes \mathscr {Y}$-measurable function. 
Assume that 
\begin{eqnarray*}
\int _Xf(x, y)^{-1}\mu (dx) < \infty , \ \ \nu \mbox {-a.e. } y
\end{eqnarray*}
and 
\begin{eqnarray*}
\int _Yf(x, y)\nu (dy) < \infty , \ \ \mu \mbox {-a.e. } x. 
\end{eqnarray*}
Then it holds that 
\begin{eqnarray*}
\int _X\left( \int _Yf(x, y)\nu (dy)\right) ^{-1}\mu (dx) \leq 
\left( \int _Y\left( \int _Xf(x, y)^{-1}\mu (dx)\right) ^{-1} \right) ^{-1}. 
\end{eqnarray*}
\end{theorem}

\begin{cor}\label{cor_harmonic}
Let $(X, \mathscr {X}, \mu )$ be a probability space and $(Y, \mathscr {Y}, \nu )$ be a measure space. 
Let $f : X\times Y\longrightarrow (0, \infty )$ be an integrable function. 
Then it holds that 
\begin{eqnarray}\label{ineq_harmonic}
\int _YH(f(\cdot , y) ; \mu )\nu (dy) \leq 
H\left( \int _Yf(\cdot , y)\nu (dy) ; \mu \right) , 
\end{eqnarray}
where $H(\varphi ; \mu ) \equiv \left( \int _X\varphi (x)^{-1}\mu (dx)\right) ^{-1}$ denotes the harmonic mean of a random variable $\varphi : X\longrightarrow (0, \infty )$. 
\end{cor}

Inequality (\ref {ineq_harmonic}) implies that ``the integral of the harmonic mean is not greater than the harmonic mean of the integral.'' 
Note that Theorem \ref {th_harmonic} and Corollary \ref {cor_harmonic} can be directly obtained by a standard argument in measure theory using the H\"older-type inequality 
(see Proposition 14.6.2 in \cite {Rao} and Exercise 9.12 in \cite {Steele}\footnote{The author thanks Prof.~Keisuke Hara for pointing it out.}). 
However, interestingly, 
we can also get (\ref {ineq_harmonic}) by using our main results when $(Y, \mathscr {Y}, \nu ) = ([0, T], \mathcal {B}([0, T]), dt)$. 
Indeed, from Theorems \ref {th_ant}--\ref {th_stat}, we have that 
\begin{eqnarray*}
\frac{\kappa }{2} + \tilde{\kappa }\E \left[ \frac{1}{V_T}\right]  = \inf _{{\boldsymbol \zeta }\in \mathcal {A}^\mathrm {ant}(1)}\E [\mathcal {C}({\boldsymbol \zeta })] \leq 
\inf _{{\boldsymbol \zeta }\in \mathcal {A}^\mathrm {stat}(1)}\E [\mathcal {C}({\boldsymbol \zeta })] = 
\frac{\kappa }{2} + \frac{\tilde{\kappa }}{U_T}, 
\end{eqnarray*}
which immediately implies that 
\begin{eqnarray*}
\int ^T_0H(v_t ; P)dt = U_T \leq \E \left [\frac{1}{V_T}\right ]^{-1} = H\left( \int ^T_0v_tdt  ; P \right) . 
\end{eqnarray*}
Note that we can easily show Theorem \ref {th_harmonic} for general $(X, \mathscr {X}, \mu )$ and $(Y, \mathscr {Y}, \nu )$ by considering similar optimization problems to our model.

\end{document}